\documentclass[conference]{IEEEtran}\usepackage[T1]{fontenc}
\usepackage[utf8]{inputenc}
\usepackage{amsmath} 
\usepackage{amsfonts} 
\usepackage{amssymb} 
\usepackage{amsthm}
\usepackage{color}
\usepackage{url}
\usepackage{caption}
\usepackage{subcaption}
\usepackage[linesnumbered,ruled,vlined]{algorithm2e}
\usepackage{setspace} 
\usepackage{verbatim}
\usepackage{multirow}
\usepackage{graphicx}
\usepackage{comment}

\newtheorem{Proposition}{Proposition}

\SetKwInput{KwInput}{Input}                
\SetKwInput{KwOutput}{Output}              
\SetKwInput{Kwinitialize}{Initialization}              

\theoremstyle{remark}

\usepackage{cite}
\renewcommand{\vec}[1]{{\bf{#1}}} 
\newcommand{\vecgreek}[1]{{\boldsymbol{#1}}} 
\newcommand{\tran}{^{\mbox{\scriptsize T}}}
\newcommand{\herm}{^{\mbox{\scriptsize H}}}

\DeclareMathOperator{\Tr}{Tr}

\newcommand{\thetab}{\vecgreek{\theta}}
\newcommand{\Lambdab}{\vecgreek{\Lambda}}
\newcommand{\lambdab}{\vecgreek{\lambda}}

\newcommand{\Sigmab}{\vec{\Sigma}}
\newcommand{\mybibliography}{\bibliography{conf_short,jour_short,Bi}}

\begin{document}
\title{User Activity Detection and Channel Estimation of Spatially Correlated Channels via AMP in Massive MTC}
\author{
\IEEEauthorblockN{Hamza Djelouat, Leatile Marata,  Markus Leinonen, Hirley Alves, and Markku Juntti}
\IEEEauthorblockA{Centre for Wireless Communications -- Radio Technologies,   University of Oulu, Finland\\
e-mail: \{hamza.djelouat,leatile.marata,markus.leinonen,hirley.alves,markku.juntti\}@oulu.fi.\\
}

}

\maketitle

\begin{abstract}
This paper addresses the problem of joint user identification and channel estimation (JUICE) for grant-free access in massive machine-type communications (mMTC). We consider the JUICE under a spatially correlated fading channel model as that reflects the main characteristics of the practical multiple-input multiple-output channels. We formulate the JUICE as a sparse recovery problem in a multiple measurement vector setup and present a solution based on the approximate message passing (AMP) algorithm that takes into account the channel spatial correlation. Using the state evolution, we provide a detailed  theoretical analysis on the activity detection performance of AMP-based JUICE by deriving  closed-from expressions for the probabilities of miss detection and false alarm. The  simulation experiments  show that the performance predicted   by the theoretical analysis matches the one obtained by the numerical results. 
\end{abstract}

\begin{IEEEkeywords}
mMTC, activity detection,  AMP, spatial correlation. 
\end{IEEEkeywords}

\section{Introduction}
Grant-free access is  as a key enabler for massive machine-type communications (mMTC)  in order to  support  connectivity to a large number of users with sporadic activity and small signalling overhead requirements \cite{cirik2019toward}. 
In contrast to the conventional scheduled access,  mMTC with  grant-free access allows the active users, referred to herein as user equipments (UEs), to directly transmit  information along with their unique signatures to the base station (BS) without prior scheduling, thus, resulting in collisions which is solved at the BS through  joint user identification and channel estimation (JUICE).

Motivated by the sparse user activity pattern, several sparse recovery algorithms have been proposed to solve the JUICE such as,  approximate message passing (AMP) \cite{chen2018sparse,liu2018massive,liu2018massiveII,senel2018grant,ke2020compressive}, sparse Bayesian learning (SBL) \cite{zhang2017novel},  mixed-norm minimization \cite{Djelouat-Leinonen-Juntti-21-icassp,Djelouat2020Joint,djelouat2021spatial}, and  deep learning \cite{cui2020jointly}. In particular, AMP has been widely investigated  in the context of JUICE for mMTC with  grant-free access. For instance, Chen {\it et al.} \cite{chen2018sparse} analyzed analytically  the user activity detection performance  for AMP-based JUICE solution in both single measurement vector (SMV) and multiple measurement vector (MMV)  setups. Liu and  Yu \cite{liu2018massive,liu2018massiveII} extended the analysis presented in \cite{chen2018sparse} and  derived  an asymptotic performance analysis for JUICE in terms of activity detection, channel estimation, and the achievable rate. Senel and Larsson \cite{senel2018grant} proposed a ``non-coherent''   detection scheme for very short packet transmission based on a modified version of AMP in order to jointly detect  the active users along with  their transmitted information bits. Ke {\it et al.} \cite{ke2020compressive}  addressed  the JUICE problem in an enhanced mobile broadband  system and proposed a generalized AMP algorithm that exploits the sparsity in  both the spatial and the angular domains. 

The vast majority of AMP-based JUICE  works assume that the multiple-input multiple-output (MIMO) channels follow an uncorrelated channel model. Although this assumption leads to derive analytically tractable solutions  \cite{chen2018sparse,liu2018massive}, it is not always practical as the MIMO  channels are almost always spatially correlated \cite{bjornson2016massive}. Therefore, the performance analysis presented in the aforementioned works may not be suitable for practical scenarios \cite{cui2020jointly}. Recently, few works addressed the JUICE in \textit{spatially correlated} MIMO channels. For instance, several mixed-norm minimization formulations using different levels of prior knowledge of the channel distribution information (CDI)  have been proposed in \cite{Djelouat2020Joint,Djelouat-Leinonen-Juntti-21-icassp,djelouat2021spatial}, whereas, Chen  {\it et al.} \cite{cheng2020orthogonal} presented an orthogonal AMP algorithm to exploit both the spatial and the temporal channel correlation in mMTC systems. While these works have investigated a more practical JUICE setup, they did not provide any theoretical analysis on the user activity detection performance for the JUICE problem.

This paper aims to provide more insights on the JUICE performance under a more practical  channel model. In particular, we utilize a Bayesian AMP algorithm to solve the JUICE in spatially correlated MIMO channels. Furthermore, the paper provides a detailed analytical study for user activity detection performance by deriving closed-form expressions for the probabilities of miss detection and false alarm. The  simulation experiments show that the predicted theoretical analysis matches the  numerical results.

\section{System Model and Problem Formulation}
\label{sec::system}
\subsection{System Model}
We consider a single-cell uplink mMTC network consisting of a set $\mathcal{N}=\{1,\ldots,N\}$  of uniformly distributed single-antenna UEs communicating with a  BS  equipped with a uniform linear array (ULA) containing $M$ antennas.  We consider a block fading channel response over each coherence period $T_{\mathrm{c}}$. Let $\vec{h}_i \in\mathbb{C}^{M}$ denotes the channel response  from the $i$th UE to the BS. We consider that the channel $\vec{h}_i$ follows a correlated Rayleigh fading channel model given as 
\begin{equation}\label{eq::ch}
    \vec{h}_i = \vec{R}_i^\frac{1}{2} \bar{\vec{h}}_i, \quad \forall i \in \mathcal{N},
\end{equation}
where $\vec{R}_i=\mathbb{E}[\vec{h}_i\vec{h}_i\herm] \in \mathbb{C}^{M \times M}$ is the channel covariance matrix and $\bar{\vec{h}}_i \sim \mathcal{CN}(\vec{0},\,\vec{I}_M)$. Note that  the vast majority of JUICE-related works consider independent Rayleigh fading  channel  with $\vec{R}_i = \beta_i\vec{I}_M$,  
which is a simplified version of \eqref{eq::ch}. However,  since the  channels are typically correlated, we consider the more practical scenario with  dense covariance matrices in order to characterize the  channel spatial correlation and the average path-loss in different spatial directions \cite{bjornson2016massive}.

The channel realizations $\vec{h}_i$, $\forall i \in \mathcal{N}$,  are assumed to be independent between different  $T_\mathrm{c}$. Furthermore, we consider UEs with low mobility, which is justified in the context of mMTC \cite{laya2013random}. Hence, we adopt the common assumption that the channels are  wide-sense stationary. Thus,  the  set of channel covariance matrices $\{\vec{R}_i\}_{i=1}^N$  vary in a slower time-scale  compared to the channel realizations \cite{bjornson2016massive}. Furthermore, $\{\vec{R}_i\}_{i=1}^N$ are assumed to be known to the  BS \cite{cheng2020orthogonal}.

In order to deploy a  grant-free  access scheme, we assume that all the UEs and the BS are synchronized. As the number of UEs in a typical mMTC network is very large, the BS  cannot assign the UEs with orthogonal pilot sequences, because it would require a pilot sequence length of the same order as the number of UEs. Therefore, the BS assigns to  the UEs non-orthogonal pilot sequences. More precisely, the BS  generates first a pool of random pilot sequences that are drawn, for instance, from  an independent  identically distributed (i.i.d.) Gaussian or an i.i.d.\  Bernoulli distribution. Then, the BS assigns to  each UE $i\in\! \mathcal{N}$  a  unit-norm pilot sequence $\vecgreek{\phi}_i \in \mathbb{C}^{\tau_{\mathrm{p}}}$. This paper considers that the pilot sequences are generated from a complex symmetric Bernoulli distribution.  This setup is motivated by the fact that: 1) pilot sequences generated from a complex symmetric Bernoulli distribution are  practical as they can be deployed using  quadratic phase shift keying (QPSK) modulation, 2) matrices drawn from a Bernoulli distribution are well suited for AMP-based support and signal  recovery \cite{senel2018grant,schniter2020simple} as we will discuss later.

Due to the sporadic nature of mMTC, only a small subset of UEs are active at each $T_\mathrm{c}$. At each  $T_\mathrm{c}$, the active UEs first transmit  their pilot sequences to the BS followed by transmitting the information data. The BS  uses the received pilot sequences to identify the  active UEs and  estimate their  channels in order to perform coherent data detection. Accordingly, the received signal associated with the transmitted pilots at the BS, denoted by $\vec{Y} \in \mathbb{C}^{\tau_{\mathrm{p}}\times M}$,  is given by
  \begin{equation}
\label{eq::Y}
 \vec{Y}=\textstyle\sum_{i=1}^{N}\gamma_i  \vecgreek{\phi}_i\vec{h}_i\tran+\vec{W},
\end{equation}
where $\vec{W}\!\sim \!\mathcal{CN}(\vec{0},\,\sigma^{2}\vec{I}_M) \in\!\mathbb{C}^{\tau_{\mathrm{p}}\times M}$ is additive white Gaussian noise, and $\gamma_i\!\in\mathbb{B}$ is the $i$th element of the binary user activity indicator vector $\vecgreek{\gamma}=[\gamma_1,\gamma_2,\ldots,\gamma_N]\tran$.

The activity indicator $\gamma_i$ is statistically modeled as a  Bernoulli random variable with ${\mathrm{Pr}(\gamma_i=1) =\epsilon}$ and ${\mathrm{Pr}(\gamma_i=0) = 1-\epsilon}$. Subsequently, we define the effective channel of the $i$th UE as  $\vec{x}_i=\gamma_i\vec{h}_i$.  Thus, effective channel $\vec{x}_i$ has a mixed Gaussian-Bernoulli distribution, given as
\begin{equation}
     p_{\vec{x}_i}=(1-\epsilon)\delta(\vec{h}_i)+\epsilon p_{\vec{h}_i}
\end{equation}
where ${p_{\vec{h}_i} \sim \mathcal{CN}(\vec{0},\vec{R}_i)}$ and $\delta(\cdot)$ is the Dirac delta function. We define the effective channel matrix as ${\vec{X}=[\vec{x}_1,\vec{x}_2,\ldots,\vec{x}_{N}] \in \mathbb{C}^{M\times N}}$ and the pilot sequence matrix as $\vec{\Phi}=[\vecgreek{\phi}_1,\vecgreek{\phi}_2,\ldots,\vecgreek{\phi}_N] \in \mathbb{C}^{\tau_\mathrm{p}\times N}$. Accordingly, we can rewrite the received pilot signals in  \eqref{eq::Y} as
\begin{equation}
     \vec{Y}= \vec{\Phi} \vec{X}\tran+ \vec{W}. 
     \label{eq::CS}
\end{equation}
  
\subsection{Problem Formulation}
The objective of JUICE is to identify the active UEs and  estimate their channel responses. Therefore, the JUICE reduces to identifying the locations of the non-zero columns of  the effective channel matrix $\vec{X}$ and estimating their coefficients. Since 
$\vec{X}\tran$ is  a \emph{row-sparse} matrix, the JUICE can be modelled as joint support and signal recovery from an MMV setup. Subsequently, the  canonical form of the JUICE can be presented as
\begin{equation}
       \min _{\vec{X} }\frac{1}{2}\|\vec{\Phi}\vec{X}\tran-\vec{Y}\|_{\mathrm{F}}^2 + \beta_1 \|\vec{X}\tran\|_{2,0},
     \label{eq::l_0}
\end{equation}
where $\|\vec{X}\tran\|_{2,0}\!=\!\sum_{i=1}^{N}1(\|\vec{x}_i\|_2\neq 0)$ is the sparsity promoting penalty and $\beta_1$  is a regularizer that controls the sparsity of the solution. Since the $\ell_0$-``norm'' minimization is an intractable combinatorial NP-hard problem, several algorithms have been proposed  to relax  \eqref{eq::l_0}. The existing recovery algorithms can be categorized into two classes depending on the required prior information on the sparse signal.  The first class includes mixed-norm minimization and greedy algorithms, where the recovery exploits only the sparse structure of the signal. The second class consists of algorithms that require prior information on the distributions of a  signal, for instance, AMP and SBL.  Such prior information often renders the second class of algorithms to have superior sparse support and signal recovery performances.

This paper exploits the assumption that the  CDI  is known to the BS and presents an AMP-based solution for the JUICE problem. In particular, we first provide a detailed description of AMP-based JUICE in spatially correlated channels. Second, we  evaluate analytically the activity detection performance and we derive closed-form expressions for the probability of miss detection and the probability of false alarm.

\section{AMP for JUICE with Spatially Correlated Channels}
\label{sec:AMP}
AMP is an iterative sparse recovery algorithm that has been  proposed originally in \cite{donoho2009message} for the general sparse recovery problem in  an SMV setup and extended to MMV setup in \cite{ziniel2012efficient}. Subsequently, AMP has been deployed to solve the JUICE in \cite{chen2018sparse,liu2018massive,liu2018massiveII,senel2018grant,ke2020compressive}.  In this paper, we adopt a Bayesian  AMP  originally proposed in  \cite{kim2011belief} for solving an MMV sparse recovery problem. This section provides a  description on the design of AMP for solving the JUICE  in spatially correlated channels.

The  AMP algorithm for sparse signal recovery from an MMV setup can be expressed by the following iterations \cite{kim2011belief}:   
\begin{equation}
\label{eq::denoise}
    \hat{\vec{x}}_i^{(t+1)}=\eta\big(\vec{Z}^{(t)\tran}\vec{\phi}_i^{*}+\hat{\vec{x}}_i^{(t)};\vec{\Sigma}^{(t)}\big),
\end{equation}
\begin{equation}\label{eq::Zupdate}
\vec{Z}^{(t+1)}=\vec{Y}-\vec{\Phi} \vec{X}^{(t+1)\tran}+ \frac{N}{\tau_\mathrm{p}}\textstyle\sum_{i=1}^N\frac{\eta^{\prime}\big(\vec{Z}^{(t)\tran}\vec{\phi}_i^{*}+\vec{x}_i^{(t)}\big)}{N}, 
\end{equation}
where $t=1,2,\ldots$ is the iteration index, $\hat{\vec{X}}^{t}=[\hat{\vec{x}}_1^{(t)},\ldots, \hat{\vec{x}}_N^{(t)}]$ is the estimate
of $\vec{X}$ at iteration $(t)$, $\vec{Z}^{(t)}\in  \mathbb{C}^{\tau_{\mathrm{p}}\times M}$ is the residual matrix initialized with $\vec{Z}^{(1)}=\vec{Y}$, and $\Sigmab^{(t)}\in \mathbb{C}^{M\times M}$ denotes a covariance matrix that can be tracked using the state evolution as we discuss later. Function  $\eta(\cdot)$ represents the denoising function that operates on each row of $\vec{X}\tran$ individually, and  $\eta^{\prime}(\cdot)$ is the first-order derivative of $\eta(\cdot)$. The third term in \eqref{eq::Zupdate} is called the
Onsager term and it is the key component in determining the performance of AMP \cite{donoho2009message}. 

For the matrix $\vec{\Phi}$ drawn from a Bernoulli distribution and under the assumption that $N,\tau_{\mathrm{p}}\rightarrow\infty$ with a fixed ratio $\frac{\tau_{\mathrm{p}}}{N}$, the term $\thetab_i^{(t)}=\vec{Z}^{(t)\tran}\vec{\phi}_i^{*}+\vec{x}_i^{(t)}$, $\forall i \in \mathcal{N}$, is statistically equivalent to the sum  of the true effective  channel  $\vec{x}_i$  and a colored noise term  $\vec{e}^{(t)}\sim\mathcal{CN}(\vec{0},\vec{\Sigma}^{(t)})$ as follows
    \begin{equation}
    \label{eq:CNsig}
        \vecgreek{\theta}_i^{(t)}=\vec{x}_i+\vec{e}^{(t)}, \quad \forall i \in \mathcal{N}.
\end{equation}

Given the linear signal model \eqref{eq:CNsig} and by exploiting  the fact that the  CDI is known to the BS, a minimum mean square error (MMSE) based denoiser function $    \eta(\vecgreek{\theta}_i^{(t)};\vec{\Sigma}^{(t)})$ is calculated as
\begin{equation}\label{eta_mmse}
\begin{array}{ll}
     \eta(\vecgreek{\theta}_i^{(t)};\vec{\Sigma}^{(t)})\!\!\!\!\!\!&=\mathbb{E}[\vec{x}_i|\thetab_i^{(t)}]\\
       &=\psi(\thetab_i^{(t)};\vec{\Sigma}^{(t)})\vec{R}_i(\vec{R}_i+\vec{\Sigma}^{(t)})^{-1}\vecgreek{\theta}_i^{(t)},  \forall i\in \mathcal{N},  
\end{array}
\end{equation}
where 
\begin{equation}\label{eq::psi}
  \psi(\vecgreek{\theta}_i^{(t)};\vec{\Sigma}^{(t)})=\Big(1+\frac{1-\epsilon}{\epsilon}  \exp\big({u_i^{(t)}-w_i^{(t)}}\big)\Big)^{-1},
\end{equation}
with 
$w_i^{(t)}=\vecgreek{\theta}_i^{(t)\herm}\vecgreek{\Xi}_i^{(t)}\vecgreek{\theta}_i^{(t)}$, $\vecgreek{\Xi}_i^{(t)}=\vec{\Sigma}^{{(t)}^{-1}}-(\vec{R}_i+\vec{\Sigma}^{(t)})^{-1}$, and $u_i^{(t)}=\log(\frac{|\vec{R}_i+\vec{\Sigma}^{(t)}|}{|\vec{\Sigma}^{(t)|}})$.

The covariance matrix of the noise term  $\vec{\Sigma}^{(t)}$  can be tracked in the asymptotic regime via the state evolution. More precisely,  
the matrix $\vec{\Sigma}^{(t)}$ is updated at each iteration $(t)$ using the following update rules \cite{kim2011belief}
\begin{equation}\label{SE}
    \begin{array}{ll}
         \hspace{-3mm}\vec{\Sigma}^{(1)}\hspace{-3mm}&=\sigma^2\vec{I}_M +\mathbb{E}[\vec{X}\vec{X}\herm] \\
          \hspace{-3mm}\vec{\Sigma}^{(t+1)}\hspace{-3mm}&=\sigma^2\vec{I}_M +\frac{N}{\tau_{\mathrm{p}}}\frac{1}{N}\sum_{i=1}^N ( \psi_i^{(t)}-\psi_i^{(t)^2})\vec{q}_i^{(t)}\vec{q}_i^{(t)\herm}\\
          &\hspace{3mm}+\psi_i^{(t)}\vec{\Sigma}^{(t)}\vec{R}_i(\vec{R}_i+\vec{\Sigma}^{(t)})^{-1},
    \end{array}
\end{equation}
where  $\psi_i^{(t)}=\psi(\thetab_i^{(t)};\vec{\Sigma}^{(t)})$, and $\vec{q}_i^{(t)}=\vec{R}_i(\vec{R}_i+\vec{\Sigma}^{(t)})^{-1}\thetab_i^{(t)}$.

\section{Activity Detection Performance}

In this  section, we derive  closed-form expressions for the probabilities of miss detection and false alarm achieved by  AMP in spatially correlated fading channels. The derivation hinges mainly on the equivalent system model \eqref{eq:CNsig} and the state evolution matrix \eqref{SE}. While the Gaussian system model in \eqref{eq:CNsig}  holds  in the  asymptotic regime, it can provide useful insight on the performance of the AMP  for the practical mMTC setup where the number of connected UEs $N$ is large, yet finite.

\subsection{Decision Threshold}
Here, we discuss the decision rule for the  user activity detector on the AMP output.  Let us examine  the denoising function $\eta(\cdot) $ given in \eqref{eta_mmse}. 

Note that for  any finite $\epsilon$,  $\psi(\thetab_i^{(t)};\vec{\Sigma}^{(t)}) \in [0,1]$. Thus, by a closer look, one can see that the denoising function $\eta(\cdot,\cdot)$ consists of an activity indicator estimate $\psi(\thetab_i^{(t)};\vec{\Sigma}^{(t)}) \in [0,1]$ and a conventional MMSE estimate term $\vec{R}_i(\vec{R}_i+\vec{\Sigma}^{(t)})^{-1}\vecgreek{\theta}_i^{(t)}$, $\forall i\in \mathcal{N}$. Therefore, in order to set the decision rule for the user activity  detector,  one can use  $\psi(\cdot\,;\,\cdot)$ to  determine the activity status of each UE. More precisely,  if $\psi_i^{(t)}\rightarrow 1$, the $i$th UE is declared active, and  if $\psi_i^{(t)}\rightarrow 0$, the $i$th UE is declared not  active. While the activity detection  performance can be characterized at each iteration $(t)$, it is typically more interesting to discuss performance upon AMP convergence. Thus, we omit the iteration index $(t)$ in the following derivations for brevity.

In a practical scenario, one would use a vector of pre-defined threshold values $\vec{l}=[l_1,\ldots,l_N]\tran$, such that  $l_i \in [0,1]$, ${\forall i \in \mathcal{N}}$. The activity detector will  declare the   $i$th UE to be active  if $\psi_i\geq l_i$, and  inactive otherwise. The values of $l_i$, $\forall i \in \mathcal{N}$, can be selected based on the cost of miss detection and the cost of false alarm for each UE.  This paper proposes the following  decision rule on the UEs activity: \begin{equation}\label{eq:thr}
 \hat{\gamma_i} \!=\begin{cases}
1,& \!\!\!\!\thetab_i\herm\vecgreek{\Xi}_i\thetab_i\!\!\geq\alpha_i= u_i-\log\big(\frac{\epsilon(1-l_i)}{l_i(1-\epsilon)}\big)\\ 
0,& \!\!\!\!\thetab_i\herm\vecgreek{\Xi}_i\thetab_i\!\!<\alpha_i=u_i-\log\big(\frac{\epsilon(1-l_i)}{l_i(1-\epsilon)}\big)
\end{cases}, \forall i \in \mathcal{N}. \!
\end{equation}

\subsection{Probabilities of Miss Detection and False Alarm}
The activity detection performance  is quantified using two types of error probabilities. First, the probability of false alarm, which represents  the probability of declaring   an inactive UE to be active. Second, the probability of miss detection, which represents the probability of declaring an active UE as inactive. 

The equivalent signal model \eqref{eq:CNsig} suggests that  the term  $\thetab_i$, $\forall i \in \mathcal{N}$,  follows a zero-mean complex Gaussian distribution, i.e., $ \thetab_i \sim \mathcal{CN}(\vec{0},\vec{C}_i)$. The   covariance matrix  $\vec{C}_i$ depends  on the value of the true  $\gamma_i$,  and it is given as
\begin{equation}\label{eq:C}
\vec{C}_i =\begin{cases}
\Sigmab+\vec{R}_i, & \gamma_i=1\\ 
\Sigmab,&\gamma_i=0 
\end{cases}, \forall i \in \mathcal{N}. 
\end{equation}

In order  characterize the activity detection performance, we refer to  \eqref{eq:thr} and  define two complex  quadratic Gaussian random variables $Q_{1,i}$ and $Q_{0,i}$, $\forall i \in \mathcal{N}$,  as 
\begin{equation}\label{eq:GQrv}
    \begin{array}{ll}
        Q_{1,i}\hspace{-3mm}&= (\thetab_i\herm\vecgreek{\Xi}_i\thetab_i|\gamma_i=1) \\
          Q_{0,i}\hspace{-3mm}&= (\thetab_i\herm\vecgreek{\Xi}\thetab_i|\gamma_i=0) 
    \end{array}, \forall i\in \mathcal{N}.
\end{equation}
Next, by using the decision rule in \eqref{eq:thr} and the 
random variables in \eqref{eq:GQrv},  the probabilities of the   miss
detection and false alarm for each UE $i$ are defined,  respectively,  as 
\begin{equation}\label{MD}
P^{\mathrm{MD}}_{i}=\mathrm{Pr}(\hat{\gamma_i}=0|\gamma_i=1)=\mathrm{Pr}(Q_{1,i}\leq\alpha_i\big), \forall i \in \mathcal{N}.
 \end{equation}
 \begin{equation}\label{FA}
P^{\mathrm{FA}}_{i}=\mathrm{Pr}(\hat{\gamma_i}=1|\gamma_i=0)=\mathrm{Pr}(Q_{0,i}>\alpha_i\big) ,  
\forall i \in \mathcal{N}.
\end{equation}

Now, let us consider a general complex Gaussian quadratic  form  ${Q=\thetab\herm\vecgreek{\Xi}\thetab}$ for a random variable ${\thetab \sim \mathcal{CN}(\vec{0},\vec{C}) \in \mathbb{C}^M}$. By using some algebraic transformations, $Q$  can be expressed as a linear combination of chi-squared random variables,  as we show next. First, we  write $\thetab$ as 
\begin{equation}\label{eq::Q}
    Q=\vec{z}\herm\vec{C}^{\frac{1}{2}\herm}\vecgreek{\Xi}\vec{C}^{\frac{1}{2}}\vec{z},
\end{equation}
where ${\thetab=\vec{C}^{\frac{1}{2}} \vec{z}}$ for  ${\vec{z} \sim \mathcal{CN}(\vec{0},\vec{I}_M)}$. Let  $\vec{U}$ and ${\Lambdab=\text{diag}(\lambda_1,\ldots,\lambda_M)}$
denote  the eigenvectors and the eigenvalues, respectively, associated with ${\vec{C}^{\frac{1}{2}\herm}\vecgreek{\Xi}\vec{C}^{\frac{1}{2}}}$. Thus, we can further express $Q$ as  
\begin{equation}
    Q=\vec{z}\herm\vec{U}\Lambdab\vec{U}\herm\vec{z}=\vec{v}\herm\Lambdab\vec{v}=\textstyle\sum_{m=1}^M\lambda_m |v_m|^2,
\end{equation}
where ${\vec{v}=[v_1,\ldots,v_M]\tran=\vec{U}\herm\vec{z}}$. Note that since $\vec{U}$ is a unitary matrix, $\vec{v}$ follows the same distribution as $\vec{z}$, i.e., $\vec{v} \sim \mathcal{CN}(\vec{0},\vec{I}_M)$. We can rewrite   $v_m=\frac{1}{\sqrt{2}}(a_m+jb_m)$ where $a_m, b_m\sim \mathcal{N}(0,1)$, for $m=1,\dots,M$. Therefore, $Q$ can be finally expressed by a  linear combination of zero-mean independent squared Gaussian random variables  as
\begin{equation}\label{eq:Q_chi2}
   Q= \textstyle\sum_{m=1}^M\frac{1}{2}\lambda_m (a_m^2+b_m^2) = \textstyle\sum_{m=1}^M\frac{1}{2}\lambda_m V_m,
   \end{equation}
where $V_m=a_m^2+b_m^2$, $m=1,\ldots,M$. Note that $Q$ can be viewed as a linear combination of $M$ independent chi-squared random variables $V_m \sim\chi_2^2(1)$  with two degrees of freedom. Subsequently, a closed-form expression for the cumulative distribution function (CDF)  of  $Q$ in \eqref{eq:Q_chi2} is given as \cite[Sect.~4.3]{mathai1992quadratic} 
 \begin{equation}\label{cdf}
    F_Q(\alpha)=\mathrm{Pr}(Q\leq\alpha)=\textstyle\sum_{m=1}^{M}g(\lambdab;m) \frac{\bar{\gamma}(1,\frac{2\alpha}{\lambda_m})}{\Gamma(1)}
\end{equation}
where 
\begin{equation}
    g(\lambdab;m)=\prod_{j=1}^{M}\Big(\frac{-1}{\lambda_j}\Big)\lambda_m\lim_{k\to\frac{1}{\lambda_m}} \bigg(\big(k-\frac{1}{\lambda_m}\big)\prod_{j=1}^{M}\big(k-\frac{1}{\lambda_j}\big)^{-1}\bigg),
\end{equation}
with $\Gamma(\cdot)$ denoting  the gamma function, and  $\bar{\gamma}(\cdot\,;\,\cdot)$ denoting  the lower incomplete gamma function. 

Now after providing a closed-form expression for the CDF  of a general complex quadratic form $Q$ given in \eqref{eq::Q}, we present Proposition~\ref{prop1} which characterizes the activity detection performance based on the state evolution matrix $\Sigmab$.
\begin{Proposition}\label{prop1} Consider user activity detection by the Bayesian AMP
for mMTC in spatially correlated channels  with finite $\frac{N}{\tau_{\mathrm{p}}}$ and  largely enough $N$ and $\tau_{\mathrm{p}}$ so that the equivalent signal model \eqref{eq:CNsig} holds. Using the  CDF expression \eqref{cdf}, the probabilities of miss detection
and false alarm per UE are given, respectively, as

\begin{equation}\label{eq::md}
    P_{i}^{\mathrm{MD}}=\textstyle\sum_{m=1}^{M}g(\hat{\lambdab}_i;m) \frac{\bar{\gamma}\big(1,2\alpha_i\hat{\lambda}_{i,m}\big)}{\Gamma(1)}
\end{equation}

\begin{equation}\label{eq::fa}
    P_{i}^{\mathrm{FA}}=1-\textstyle\sum_{m=1}^{M}g(\bar{\lambdab}_i;m) \frac{\gamma\big(1,2\alpha_i/\bar{\lambda}_{i,m}\big)}{\Gamma(1)},
\end{equation}
where $\hat{\lambdab}_i=[\hat{\lambda}_{i,1},\ldots,\hat{\lambda}_{i,M}]\tran$ and $\bar{\lambdab}_i=[\bar{\lambda}_{i,1},\ldots,\bar{\lambda}_{i,M}]\tran$ are the eigenvalues associated with  
${(\vec{R}_i+\Sigmab)^{\frac{1}{2}\herm}\vecgreek{\Xi}_i(\vec{R}_i+\Sigmab)^{\frac{1}{2}}}$ and $\Sigmab^{\frac{1}{2}\herm}\vecgreek{\Xi}_i\Sigmab^{\frac{1}{2}}$, respectively.

\end{Proposition}

\section{Simulation Results}
We consider a single cell of  radius of $100$~m with a BS equipped with $M$ antennas and surrounded by $N=1000$ uniformly distributed single-antenna UEs. We consider an activity level ${\epsilon=0.05}$ at each  $T_\mathrm{c}$.  The channel responses $\vec{h}_i$, for ${i \in \mathcal{N}}$,   are generated using a local scattering  model for the channel covariance matrices \cite[Sect.~2.2]{massivemimobook}. To mitigate the channel gain differences between the UEs, we assume that each UE has a unit average channel gain, i.e., $\frac{1}{M}\Tr(\vec{R}_i)=1$, $\forall i \in \mathcal{N}$. Furthermore, each UE is assigned with a  normalized QPSK sequence $\vecgreek{\phi}_i$  drawn from an i.i.d. complex Bernoulli distribution.

\begin{figure*}[t]
    \centering
     \begin{subfigure}{0.49\textwidth}  \centering
    \includegraphics[height=5cm, width=8cm]{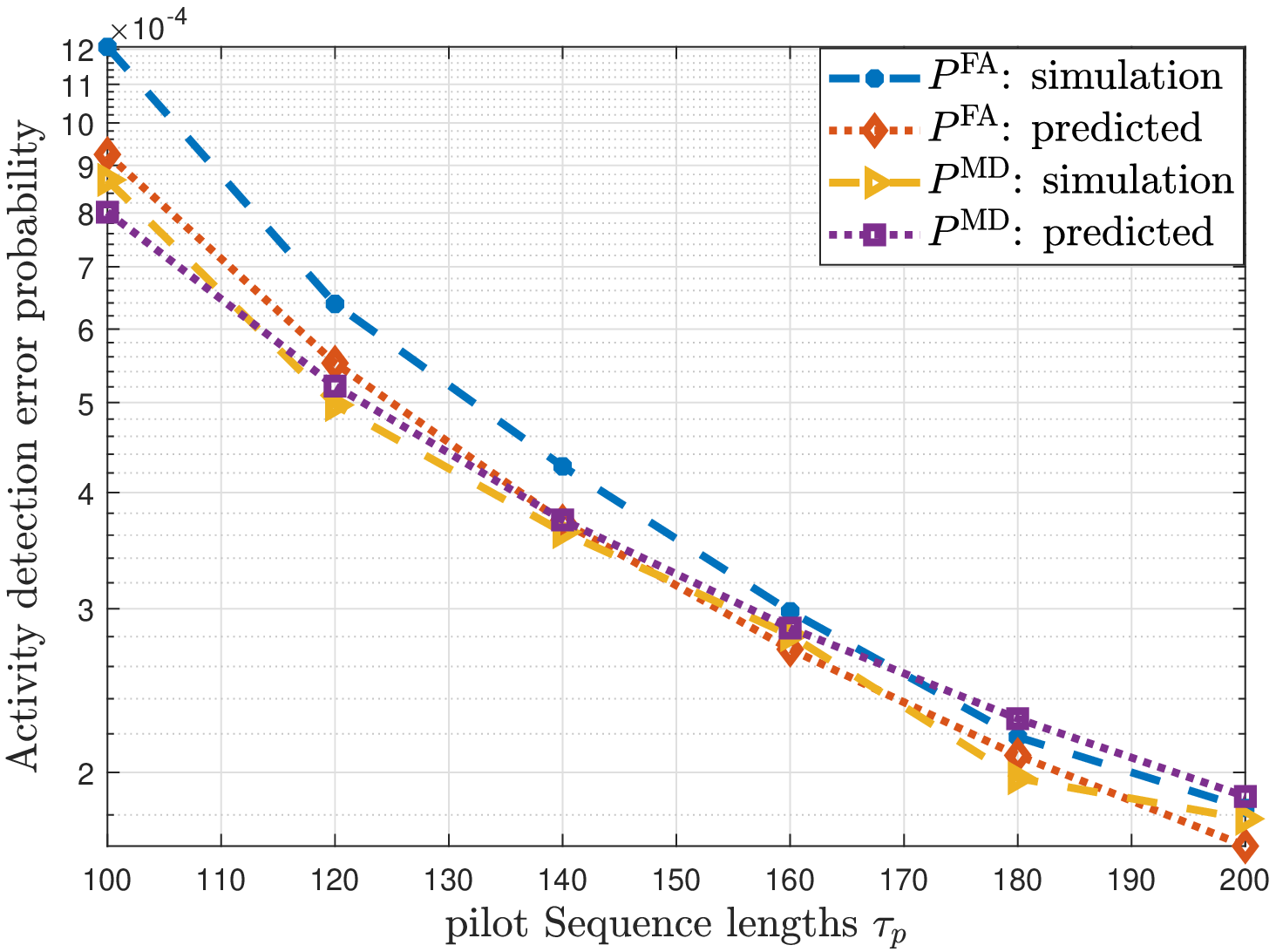}
    \caption{}
    \label{fig:Pe_pilot}
\end{subfigure}
 \begin{subfigure}{0.49\textwidth}
  \centering\vspace{.2cm}
     \includegraphics[height=5cm, width=8cm]{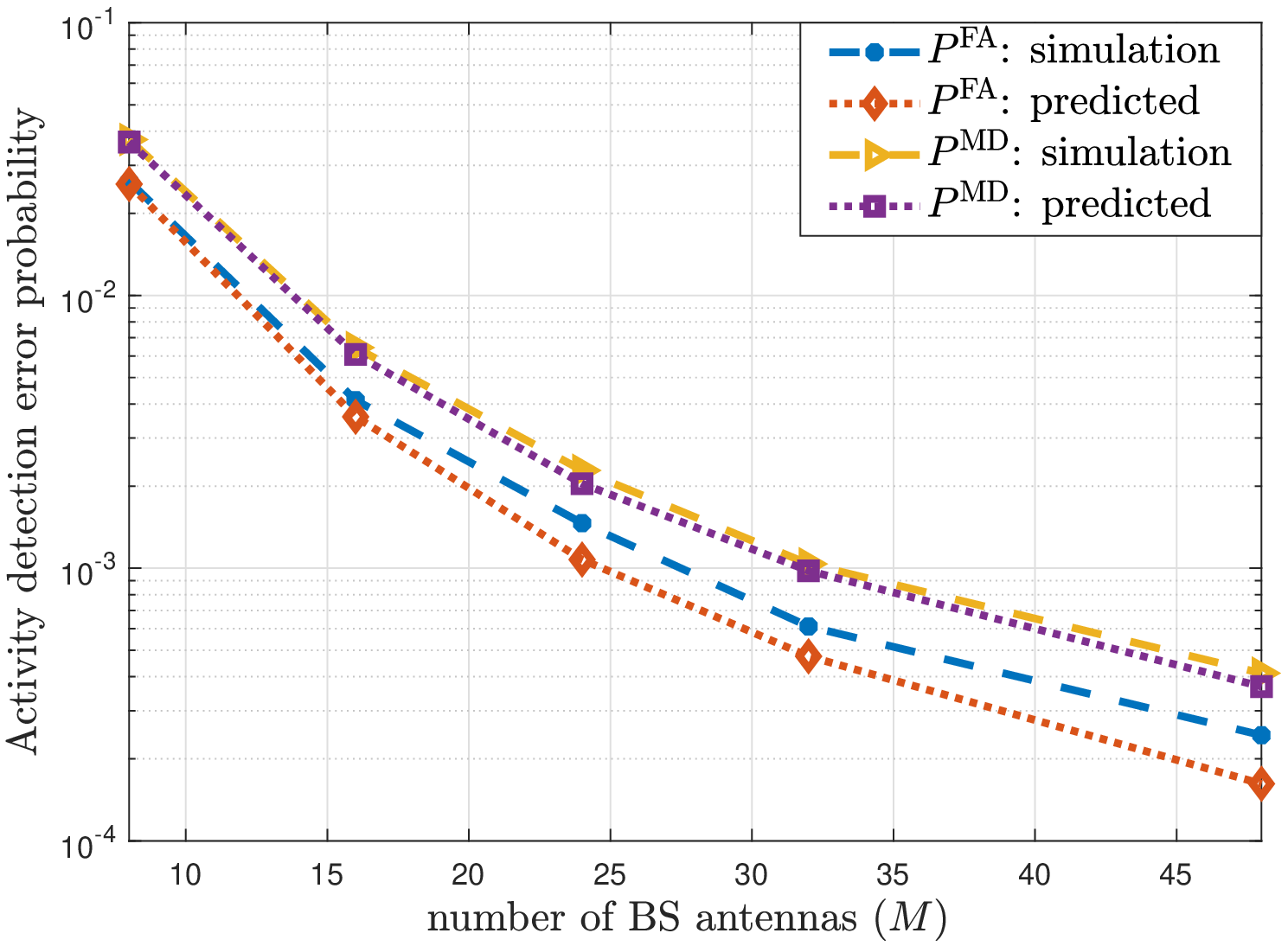}
     \caption{}
     \label{fig:Pe_BS}
\end{subfigure}  
\vspace{-3mm}
\caption{
Probabilities of miss detection and false alarm obtained by the AMP algorithm (simulation) and the derived theoretical results (predicted)  versus: a) pilot length $\tau_{\mathrm{p}}$, b) number of BS antennas $M$.
}
\label{fig:activity_detection}
\vspace{-5mm}
\end{figure*}

The activity detection performance is quantified in terms of the average miss detection ${P^{\mathrm{MD}}=\textstyle\frac{1}{N}\sum_{i=1}^NP_{i}^{\mathrm{MD}}}$ and false alarm  ${P^{\mathrm{FA}}=\textstyle\frac{1}{N}\sum_{i=1}^NP_{i}^{\mathrm{FA}}}$ probabilities.  
Fig.\ \ref{fig:activity_detection} compares the the simulated and the predicted $P^{\mathrm{MD}}$ and $P^{\mathrm{FA}}$. The simulated performance is obtained  by  running the AMP algorithm and deploying the decision rule \eqref{eq:thr} to detect the active UEs, whereas the predicted performance is obtained from the analytical expressions \eqref{eq::md} and \eqref{eq::fa}. Fig.\ \ref{fig:activity_detection}(a) shows the activity detection performance versus the pilot sequence length  $\tau_{\mathrm{p}}$ with  ${M=32}$ and a signal-to-noise ratio ${\text{(SNR)}=10}$~dB. The obtained results suggest that  increasing $\tau_{\mathrm{p}}$ results in continual improvement in the activity detection performance.  More interestingly, the results show clearly that the probabilities of miss detection and false alarm derived in \eqref{eq::md} and \eqref{eq::fa}  match  the simulation results obtained by  AMP. Fig. \ref{fig:activity_detection}(b) shows the same activity detection performance but  with respect to the number of  BS antennas $M$. As expected, the activity detection performance improves significantly with  increasing $M$. Furthermore, similar to the results  in Fig.  \ref{fig:activity_detection}(a), the simulated results provide the same performance as the theoretical ones.   

\begin{figure}
    \centering
    \includegraphics[height=4.7cm, width=8cm]{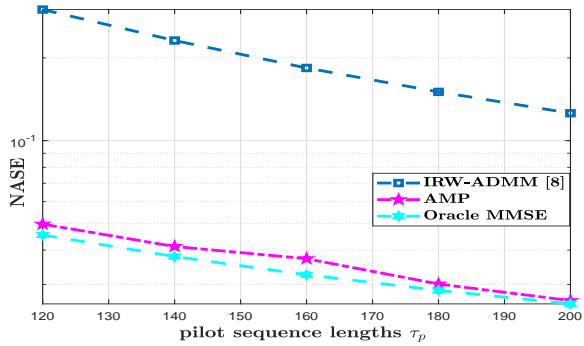}
    \caption{Channel estimation performance in terms of NASE versus pilot length $\tau_{\mathrm{p}}$  for $M=32$, and ${\text{SNR}=10}$~dB.}
    \label{fig:NASE}\vspace{-4mm}
\end{figure}

Fig.\ \ref{fig:NASE} illustrates the channel estimation performance in terms of  normalized average square error (NASE), defined as $\textstyle\frac{\mathbb{E}\left [\Vert \vec{X}_\mathcal{S}-\hat{\vec{X}}_\mathcal{S}\Vert_{\mathrm{F}}^2 \right ]}{\mathbb{E}\left[\Vert \vec{X}_\mathcal{S}\Vert_{\mathrm{F}}^2 \right]}$, 
where  the expectation is computed via Monte-Carlo averaging over all sources of randomness and $\mathcal{S}$ denotes the set of true active UEs. Fig.\ \ref{fig:NASE} compares  AMP with two algorithms: 1) the oracle MMSE estimator, which is provided ``oracle'' knowledge on $\mathcal{S}$, and 2) IRW-ADMM proposed in \cite{Djelouat-Leinonen-Juntti-21-icassp}, where the JUICE is formulated as an iterative reweighted $\ell_{2,1}$-norm minimization  and solved via the alternating direction method of multipliers. Fig.\ \ref{fig:NASE} shows that AMP   outperforms significantly IRW-ADMM   and it provides near-optimal NASE performance as it approaches the lower bound obtained by the oracle MMSE estimator. This result highlights the gain obtained by using rich side information at the BS: while IRW-ADMM operates only on the mere fact that the effective channel is sparse, AMP utilizes  both  channel and  noise statistics, leading to  better channel estimation performance.

\section{Conclusion}
\label{conclusion}

This paper presented an AMP-based solution for the JUICE problem in mMTC under spatially correlated fading channels. By utilizing the properties of the state evolution of the AMP algorithm, the paper derived closed-form expression for 
both the miss detection and false alarm probabilities. The simulation experiments showed that the theoretical analysis provided in this paper matched the numerical results.

\section*{Acknowledgements}
This work has been financially supported in part by the Academy of Finland (grant 319485) and Academy of Finland 6Genesis Flagship (grant 318927). The work of M. Leinonen has also been financially supported in part by Infotech Oulu and the Academy of Finland (grant 340171 and 323698). H. Djelouat  would like to acknowledge the support of Tauno Tönning Foundation,  Riitta ja Jorma J. Takanen Foundation, and Nokia Foundation. L. Marata's work is supported partially by Botswana International University of Science and Technology.
\bibliographystyle{IEEEtran}
\mybibliography

\end{document}